# Launching Your VR Neuroscience Laboratory

Ying Choon Wu, Christopher Maymon, Jonathon Paden, and Weichen Liu

## Contents

**Abstract** The proliferation and refinement of affordable virtual reality (VR) technologies and wearable sensors have opened new frontiers in cognitive and behavioral neuroscience. This chapter offers a broad overview of VR for anyone interested in leveraging it as a research tool. In the first section, it examines the fundamental functionalities of VR and outlines important considerations that inform the development of immersive content that stimulates the senses. In the second section, the focus of the discussion shifts to the implementation of VR in the context of the neuroscience lab. Practical advice is offered on adapting commercial, off-the-shelf devices to a researcher's specific purposes. Further, methods are explored for recording, synchronizing, and fusing heterogeneous forms of data obtained through the VR system or add-on sensors, as well as for labeling events and capturing game play. The reader should come away with an understanding of fundamental

Y. C. Wu (✉), J. Paden, and W. Liu
University of California San Diego, San Diego, CA, USA

C. Maymon
Victoria University of Wellington, Wellington, New Zealand

considerations that need to be addressed in order to launch a successful VR neuroscience research program.

# 1 Introduction

Creating a research platform that leverages the capabilities of virtual reality (VR) demands diverse expertise to a greater degree than most typical scientific endeavors centered on the human mind, brain, and behavior. While at first glance, the technical challenges of such an undertaking may seem a barrier to entry, they also create opportunities for cross-disciplinary collaboration that can drive innovation. Indeed, VR platforms can support seamless merging of experimental research with therapy, art, gaming, educational tools, and other applications. A VR-based installation, for example, can be designed to serve as both an art piece and a material for a study on esthetic experience. For this reason, one should give special advance consideration to the audience that is the target of one's project. The broad appeal and functionality of VR can readily transcend the boundaries of the specific scientific community that may be the most obvious target of one's research question; and the impact of one's study can be substantially enhanced by availing oneself of a broader range of stakeholders and strategic partnerships.

This chapter reviews key features of diverse VR systems that impact comfort and useability. It also examines several factors that are important for immersion and the cultivation of presence, which is the feeling of being physically located in a virtual space (Barbot and Kaufman 2020). While commercial game developers may seek to evoke a strong sense of presence that increases the immersive draw of a given game, researchers should consider what aspects of presence are most important for their objectives. In a study of acrophobia, for instance, it may be desirable to manipulate many sensory elements of the virtual experience – including the sounds, sights, and sensations associated with heights – in order to successfully elicit a fear of heights. On the other hand, to study aspects of spatial navigation, a virtual environment that simply replicates a sense of depth and optic flow may be sufficient. This chapter offers an overview of immersive technologies for VR – how they work and how they can be implemented in a research setting. Further, diverse approaches are examined to creating immersive content ranging from replications of real-world spaces to fantastic fictional worlds.

An additional set of vital considerations that all VR researchers must address is the types of data that they intend to record and how they plan to analyze them. Many VR systems can be either purchased off the shelf with integrated capabilities or are compatible with after-market sensors that detect and record many channels of information about a user's behavior in the game world, including head position and orientation, head movements, hand movements, gaze points, and more. Moreover, full body motion capture of the limbs and torso can be achieved with auxiliary trackers. Further, with ongoing advances in wireless and wearable biosensing technologies and signal processing techniques, it is now feasible to measure heterogeneous activities of the central, peripheral, and autonomic nervous systems in tandem with data acquired through the VR system. Indeed, the ability to integrate virtual experience with these peripheral recording systems makes VR particularly valuable in behavioral neuroscience research. However, these exciting advances also bring many challenges. This chapter offers practical advice on building an optimal research platform, acquiring and synchronizing heterogeneous modalities of data, labeling events, and capturing game play. The reader is invited to explore subsequent chapters of this volume to learn more about the state of the art in neuroimaging (chapter "Monitoring Brain Activity in VR: EEG and Neuroimaging") and eye-tracking (chapter "Eye-Tracking in VR") in VR-based research paradigms.

## 2 Behind the Magic of Virtual Reality

### 2.1 Visuo-Spatial Cues

Perhaps the most well-known approach to achieving an immersive, 360° VR experience in three dimensions is through a head-mounted display (HMD) – which is essentially a goggle-style display system that fits over the orbital region of the face and is held in place via head straps. Although all visual input is presented binocularly via a flat display system that rests only a few inches from the user's eyes, depth cues are simulated through stereopsis – that is, slightly offset perspectives of the same image are presented to each eye, usually through separate video displays or the same display with dual feeds. Fresnel plastic lenses serve to support more comfortable viewing conditions and cause images to appear situated further away than their actual position on the display.

Positional and orientational tracking allows the HMD system to update visual input as one moves through space and looks in different directions, affording the illusion of being situated in the virtual setting. Positional tracking can be accomplished by a variety of methods. Some systems (like the HTC Vive (https://www.vive.com/)) use "outside in" tracking, which involves the use of base stations positioned at opposite corners of the VR area emitting infrared (IR) lasers that are detected by photosensors on the headset and handheld controllers. By using the timing of photosensor activation, the system obtains positional estimations and can adjust the viewing angle of the virtual camera, rendering an image to each lens while accounting for the user's motion. On the other hand, standalone devices, such as the Meta Quest (https://www.meta.com/) series, use an "inside-out" system. Here, outward-facing cameras are located on the headset and generate a real-time map of the physical environment. Using this map, the headset continuously estimates the position of the user within the physical environment using computer vision algorithms like feature detection. These "inside-out" positional estimations are then combined with parallel streams of angular velocity and acceleration from inertial measurement units (IMUs) in the headset and controllers, contributing to the ongoing estimation of the headset's rotation and position.

Cave Automatic Virtual Environment (CAVE) systems (Manjrekar et al. 2014) rely on the same basic mechanisms as an HMD, such as stereopsis and positional tracking, but use displays scaled to the size of a room. The walls, ceiling, and even the floor of a CAVE are lined with flat panel displays, which can be seamlessly adjoined when bezel-free screens are used. Passive, anaglyph interlacing, or active shutter approaches are employed to achieve a stereoscopic 3D effect. In both cases, the user wears specialized glasses. Positional tracking can be achieved through sensors mounted to the glasses or attached elsewhere to the user's body. Notably, CAVEs can support multiple simultaneous users, but the environment can only update relative to a single user's position. More information on these and other vision science topics can be found in chapter "VR for Vision Science" of this volume.

> **Factors Impacting the Quality of an HMD-Based Visual Experience**
> **Screen Resolution**
> Because the visual display is viewed at very close proximity to the user's eye, low screen resolution will cause images to appear pixelated. At least 4 K resolution (3,840 × 2,160) is recommended, but successful VR-based research has been carried out at lower resolutions.
>
> **Field of View (FOV)**
> FOV is measured in degrees and represents the range of ocular visibility. The binocular human FOV approaches 180°, with around 120° overlapping for stereoscopic vision (Read 2021). Popular VR headsets such as the Meta Quest and the HTC Vive currently support FOVs between 90° and 110°. Although an HMD with a wider FOV can approximate true-to-life viewing conditions more accurately, there are also trade-offs with resolution – as a wider FOV will yield a lower pixel density.
>
> **Interpupillary Distance (IPD)**
> To avoid eye strain, it is important that the optical center of the HMD lenses aligns with interpupillary distance. In some headsets, the user's IPD is measured through integrated infrared eye cameras, and a calibration routine assists the user in attaining optimal lens spacing using an adjustment knob. Other headsets allow you to manually shift the lenses into different spacing settings based on IPD ranges, and users are expected to ascertain their IPD on their own.
>
> **Comfort and Fit**
> A poorly fitting HMD can result in shifting or sliding of the headset, muscle strain, light bleed, and other distractions and discomforts. Most HMDs are

> secured through a combination of lateral and coronal head straps or rigid plastic bands that can be adjusted to different head sizes via Velcro or an adjustment wheel. After-market accessories, such as masks or attachable covers, can improve the effectiveness and hygienics of the facial interface, which is the soft foam lining around the edges of the open portion of the headset that presses against the face.
>
> **Users Who Wear Glasses**
>
> While contact lenses are the preferred method of vision correction while wearing an HMD, glasses can be accommodated in various ways. In some headsets, such as versions of the Vive, the distance between the lens and the face can be lengthened via side knobs so that the headset can fit over most eye-wear frames. In other cases, spacers may be added between the facial interface and the headset, effectively extending the distance between the lenses and the face. Prescription lens inserts may also be purchased for many popular brands of headsets, obviating the need to wear glasses.

## 2.2 Engaging with Single- and Multi-Player Virtual Spaces

In most HMD VR systems, the perimeter of a play area – that is the physical space in which users engage in the virtual world – must be defined at the outset. When a user approaches the edges of the play area, a boundary warning grid can appear, and stepping outside of this boundary can result in exiting the virtual world. In cases when the extent of a virtual world exceeds the boundaries of a play area, a number of different techniques exist to support locomotion and other forms of movement and travel. Examples of solutions that involve minimal degrees of freedom include riding in a vehicle on a track or entering a portal, which can directly transport users from one fixed location to another. On the other hand, controllers can be used to achieve self-directed locomotion over large regions of space. Teleporting, for instance, usually involves aiming a controller or simply looking at a desired location in the scene and clicking a button to advance one's position to that location. Teleporting is often combined with walking and allows users to traverse much greater distances more quickly than would be possible through walking alone. Controllers can also be used to achieve various forms of world pulling, such as skiing, rock climbing, swimming, and ladder climbing, and to steer virtual vehicles. Finally, VR-integrated omni-treadmills support continuous naturalistic walking without controllers, and new vehicle simulation devices (like the NOVA by Eight360 (https://www.eight360.com/)) support synchronous 360° rotation while operating a vehicle in VR.

Another category of actions supported in VR is the manipulation of virtual objects and substances. Many popular systems, such as Meta Quest and Vive, can track movements and configurations of users' hands in real time, supporting naturalistic interactions with a virtual environment based on stereotyped hand gestures.

Alternatively, hand controllers can serve as a proxy for users' hands. Continuously tracking the controllers' position and rotation and mapping gestures to available buttons/levers allows users to interact with virtual objects in many ways, such as holding, moving, dropping, throwing, and so forth. In situations where physical space may be limited or when virtual objects are distant, it is also possible to program hand controllers to select and pull objects through space toward. Interactions like these can even be driven by users' gaze, which can be estimated on the basis of head orientation via a ray projected from the center of the headset. When the ray intersects with the collider of a virtual object, that object may be "selected." Additionally, Unity and Unreal game engines, both of which are software frameworks that support video game development, offer libraries that support recognition of spoken or written keywords and phrases, allowing users to effect changes in their environments or on non-player characters through simple commands (e.g., uttering "Lights on," or "Sleep").

Although the head and the hands are the primary effectors in most VR experiences, full body tracking is also feasible and is currently used in applications such as virtual martial arts. Typically, wireless sensors are strapped to the limbs and torso to obtain continuous estimates of joint position and rotation, from which a whole-body model can be computed via forward and inverse kinematics and updated in real time. Notably, a representation of the user's body can be visible to the user as well as to other users in multi-player contexts. VR technology has reached a point where multiple people can interact in the same virtual environment, whether they are physically situated in the same space or not. This feature makes VR a desirable space to investigate social interactions, as it allows researchers to manipulate variables in ecologically valid ways, without sacrificing experimental control (Pan and Hamilton 2018). In an early study, Bailenson et al. (2003) immersed participants in a virtual environment with an avatar, varying the avatar's gender, gaze behavior, and whether it was controlled by another human or by a computer. The researchers measured the distance participants maintained from these avatars when the avatar approached the participant. Behavioral measures like these may constitute more ecologically valid evidence relative to self-report measures about the extent to which participants attributed sentience to the avatar.

Analogously, Altspace VR, a social platform supporting live virtual events and gatherings (McVeigh-Schultz et al. 2018), has been used in research conducted by the first author and colleagues examining compassion fatigue in health care workers (Wu et al. in prep). The study involved meditation and compassion cultivation activities held in a virtual courtyard and studio, followed by a simulated clinic session with a "patient" (whose avatar was animated by an actor joining the virtual session from a different location). Electroencephalographic (EEG) and electrocardiographic (ECG) data were recorded, as well as behavioral measures reflecting participants' subjective empathy and compassion. This work was part of a larger longitudinal study examining the impact of virtual meditation booster sessions on compassion cultivation training.

## 2.3 Sensory Engagement and Immersion

VR represents a seductive yoking of paradoxical capabilities. On the one hand, it offers ever increasing realism in its appeal to sensory experience. On the other, it offers the unique opportunity to divorce the senses from reality through forays into the fantastic and impossible. On the realism side of this dichotomy, it is now possible to feel simulated gun shots, desert winds, or rain drops and to smell burning tire rubber or brewing coffee while gaming. On the fantastic side, it is possible to create novel experiences by changing the physics governing different forms of sensory input. A player can experience the capability of changing properties of ambient sounds through body movement, for instance, or can try out the perspective afforded by a viewpoint that is only 5 in. from the ground – or 50 feet in the air. It is also possible to isolate unitary modalities of sensory experience. For instance, listening with eyes closed to a soundscape (Schafer 1993) of a tropical rainforest at dawn via 360° audio recording can stimulate mental imagery and allow listeners to become immersed in their imagined environments in the absence of any other sensory input that would normally occur if one were actually present in the represented setting.

Current solutions to haptic and olfactory stimulation can be grouped roughly into portable, wearable, and contact-free approaches. For instance, a portable solution to creating simple force cues can be accomplished by programming the hand controllers to vibrate upon colliding with certain types of virtual surfaces. Other examples include using some of the basic outfitting of an oxygen bar to deliver different scents or air temperature and humidity sensations directly to the nostrils, simulating changes in the ambient environment. Additionally, long or short bursts of forced air directed toward the face via mounted nozzles can simulate different patterns of airflow (Rietzler et al. 2017).

Wearables offer more elaborate solutions. Gloves, sleeves, vests, masks, and even full body suits can deliver complex sensory stimulation in synchronization with specific virtual events. These devices are capable of simulating a variety of touch and force sensations, as well as feelings of heat, cold, wind, water, and sound through vibratory motors, haptic transducers, micro heaters and coolers, and even direct electrostimulation – e.g., using Teslasuits (Caserman et al. 2021). Multi-sensory masks that mesh with popular VR headsets are able not only to supply haptic stimulation to the face, but also hundreds of distinct smells. It is worth noting that some of these sensory feedback systems only interface with specific gaming or training applications, whereas others are compatible with major game engines and offer Application Programming Interfaces (APIs) and Software Development Kits (SDKs) to support development and customization.

As an alternative to wearables or portable devices, tactile sensations can be induced through mid-air haptic technologies, which modulate volleys of ultrasonic energy in order to produce discernible pressure on the surface of the skin when in the vicinity of the device. Currently, these systems work best with the hands and are capable of simulating different sensations of texture and motion, such as the feeling of trickling water or rising bubbles. These devices are intended for use in a variety of settings, including interactive kiosks; however, developers have successfully integrated mid-air haptics into VR applications.

Whereas synthesizing olfactory, haptic, and other tactile sensations often requires some type of add-on gear, immersive sound experiences can be realized through spatial audio functions that are commonly available to the VR development community and supported by most HMD systems. Spatial audio can be understood as a rendering process such that a sound source is locked to a point in space rather than a person's head (Moore 1983). Thus, even though players may be wearing headphones attached to their HMD, sounds will become quieter as they move away from its source, and louder as they draw closer. In addition, ongoing work on path-traced acoustics allows developers to model changes in a sound's acoustic energy depending on properties such as the size, shape, and materials of the virtual environment (Beig et al. 2019). Thus, a narrow, enclosed space with many reflective surfaces will yield different acoustic qualities than an open outdoor space or one filled with sound-absorbing materials such as carpets and drapes. Likewise, a sound coming from around a corner will have different properties from one coming from a direct, un-occluded source.

Notably, as an alternative to simulating various kinds of sensory feedback produced by real-world stimuli, it is also possible to integrate real-world stimuli as physical props within a virtual experience. For instance, in recent work by the second author (Maymon et al. 2023), participants balanced across a real wooden plank which corresponded to a matching VR plank appearing to extend precariously over a city street from a height of 80 stories. Similarly, another research group added real textured surfaces within their play area in order to enhance the feeling that participants were inside a virtual cave (Kisker et al. 2021).

## 2.4 Creating Immersive Content

Creating your own virtual content necessitates selecting the most suitable game engine or platform to support development. At the time of writing this chapter, Unity (https://unity.com/) and Unreal Engine (https://www.unrealengine.com/) are the most broadly used game engines, each delivering advanced, out-of-the-box solutions. However, alternatives also exist, such as Vizard (https://www.worldviz.com/vizard-virtual-reality-software), which offers analytical tools that are specifically designed to support research, and A-Frame (Neelakantam and Pant 2017), which is a web-based framework. Additionally, live, mixed reality platforms offer a framework tailored to creating social VR-based paradigms.

It is a widely held view that Unity is a better choice for novice programmers. Unity supports scripting in C#, which presents a lower barrier to entry than programming in Unreal, which uses C++. Further, because Unity has attracted a large, active community of users, as well as numerous industry partners, abundant documentation and instances of prior work are available. Unity also offers a fairly intuitive user interface, along with team collaboration tools. On the other hand, the Unreal game engine is known for more powerful rendering and much higher resolution graphics. It supports multi-player stacks and servers natively, whereas Unity can require peripheral libraries for multiple players. Unreal also offers node-based visual programming (Blueprints). It is worth noting that for both Unity and Unreal, pre-existing assets – that is, digital objects, sounds, images, animations, and other content that comprise the virtual world – as well as code, plugins, and other functionality, can all be obtained either for free or at a cost. In this way, many basic environments can be created simply by placing assets and specifying their properties in the editor.

When designing a VR study, the choice of which platform to use depends on the content that one wishes to create or use. Many different approaches requiring varying degrees of customization and development can be adopted. Obviously, presenting 360° images or videos from an existing library requires less specialized functionality than creating a novel, believable, and fully interactive world. In the first case, the visual content is captured from the real world using a 360° camera, whereas the latter case is much more complex. Customized assets must be created and programmed with the interactive features that support the desired experience. For instance, if one wishes for some objects to clatter and bounce when they are dropped, these properties must be programmatically associated with those objects. On the other hand, it is also possible to use procedural assets generated natively by a game engine or draw from existing asset libraries or packages. This approach requires considerably less investment of resources than creating custom assets from scratch. For those whose research needs require the development of a customized experience from scratch, it is advisable to avail oneself of tools that support modeling, animating, and texturing, such as Blender (https://www.blender.org/), Maya (https://www.autodesk.com/products/maya/), Houdini (https://www.sidefx.com/products), Tilt Brush (https://www.tiltbrush.com/), and many others.

While it is often desirable to construct new experimental content that is customized to a specific research question, creating optimized and visually immersive virtual environments may require a dedicated team of creators – which may be prohibitively expensive when setting up a new study. One alternative approach that researchers may wish to consider is to "mod" an existing VR game. Modding is a process whereby users can write their own code and embed them into the file structure of a particular game. The result is a version of that game that is modified by that code. Modding is available only for those simulations whose production studio releases development files (e.g., Minecraft, Skyrim). If, however, a researcher can find an existing simulation that includes a scenario that could be minimally altered to be suitable for experimentation (perhaps you want to remove on-screen distractions like the character's health bar or add in a particular recording of your experiment instructions), then one can save considerable time and budget by developing the code necessary to shape an existing simulation, rather than to create a new virtual environment from scratch.

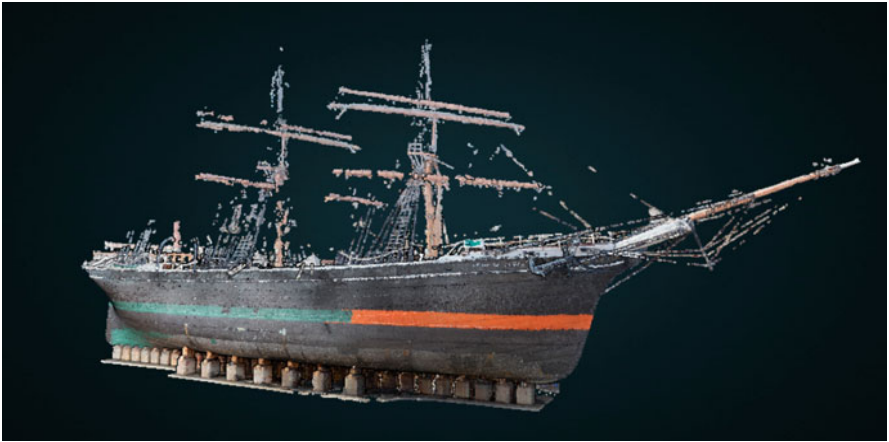

**Fig. 1** Point cloud representation of the Star of India obtained through photogrammetry (courtesy of Scott McAvoy on Open Heritage 3D)

In addition to creating assets using 3d software, it is also possible to leverage 3d capture technologies that draw upon the real world as input (Fig. 1). *Photogrammetry* is a process that involves constructing virtual 3d objects that users can walk around and view from all angles using simple photographs as inputs. This approach has become widely adopted for building at all scales – from scanning small handheld objects all the way up to large geo-regions, including cities.

Photogrammetry largely runs on *structure from motion* (SfM) techniques (Özyeşil et al. 2017), which involve estimating 3d structure from a sequence of 2d images (photographs or video). They have become a standard in software such as Agisoft Metashape (https://www.agisoft.com/), Reality Capture (https://www.capturingreality.com/), and VisualSFM (C. Wu 2011) – to name a few – and are common practice in diverse fields, including forensics and cyber archeology. The output of SfM algorithms can be stored in many forms compatible with VR practices, though the two most typical formats are LAS/LAZ point clouds or meshes. Meshes often required post processing in some form of traditional 3D software such as Blender or Maya to clean up floating points, fix holes, or modify texture maps.

An alternative approach to real-world capture is LiDAR scanning (Light Detection and Ranging), which is a laser-pulse sensing technology that can map the topography of objects and the environment. It yields point cloud representations that can be imported into modeling platforms, such as VR Sketch, so that textures and other features can be added before placement in a virtual scene. Because LiDAR lasers can penetrate foliage, it is heavily favored in contexts in which vegetation can obscure the true contours of an area (Chase et al. 2017).

## 3   VR as a Research Tool

On the one hand, VR is an attractive research tool that promises an unprecedented degree of controlled stimulation coupled with the possibility of capturing an exceptionally complex, multimodal matrix of continuous data representing diverse, temporally synchronized aspects of behavior and physiology. HMD systems track the positions of the controllers and the player's head in the virtual environment, as well as head orientation. Additional sensors added to the body and hands (e.g., haptic gloves) can provide a finely detailed, dynamic model of limb movement. Further, systems with eye-tracking technologies and ray-casting algorithms can offer continuous estimates of the user's gaze in the virtual environment.

In addition to the integrated sensors that come built-in to many VR systems, it is also possible to leverage wearable biosensing technologies that measure EEG, ECG, respiration, temperature, photoplethysmography (PPG), and more. Many of these devices are sufficiently lightweight and unobtrusive to be worn concomitantly with an HMD – and in this way, a wide range of actions and events, such as looking toward a particular location, walking, teleporting, and picking up or coming into contact with objects can be parsed and analyzed in conjunction with simultaneously recorded physiological data that reflect modulations of activities in the central and peripheral nervous systems. MRI-compatible VR systems have been developed as well (Adamovich et al. 2009).

On the other hand, as a corollary to these exciting possibilities are several hurdles that a VR-based research paradigm must overcome above and beyond the issues of game development and user experience covered in the previous section. Most commercially available VR systems do not come research-ready out of the box. To obtain access to the desired data, a license may be required or customized API or driver. Further, precautions are necessary to ensure participant safety and comfort, and to ensure that the various sensors used in a study do not interfere with one another. Groundwork must also be laid to synchronize heterogeneous modalities of data that are collected – either in real time or offline. Finally, the problem of analyzing data obtained during engagement in immersive VR is non-trivial. The past decades of cognitive neuroscience research have largely been devoted to paradigms that involve motionless volunteers viewing 2D displays and pressing buttons in response to experimentally relevant events. However, when recording EEG, ECG, or other time-series modalities that are distorted by body movement, strategies for removing or correcting motion artifacts are crucial for studies using immersive VR. Analogously, approaches for parsing saccades and fixations under conditions of head and body movement in 3D space are critical for eye-tracking work performed in immersive contexts. More information on EEG and eye-tracking in VR can be found in chapters "Monitoring Brain Activity in VR: EEG and Neuroimaging" and "Eye-Tracking in VR", respectively, of this volume.

## 3.1 Safety and Comfort in VR

A few basic precautions are advised in order to reduce the risk of injury or discomfort. First and foremost, in the case of room-scale activities, the play space should be clear of hazards that could cause users to trip, fall, or otherwise injure themselves. While wearing a headset, users should be monitored to ensure that they do not become entangled in cables or collide with walls or other stationary objects on the perimeter of the play space. Secondly, steps should be taken to ensure proper fit of the headset and adjustment of the IPD in order to avoid blurry images, ocular problems, and uncomfortable shifting or sliding of the headset during movement. Thirdly, common negative "side effects" associated with VR usage, such as motion sickness or headache, may be mitigated to some degree by the design of the virtual environment (e.g., by minimizing exposure to intense light or reducing optic flow) (Nichols and Patel 2002). Further, limiting the length of exposure (15 min is a conservative time window) and ensuring adequate breaks are important measures to minimizing negative experiences (Kaimara et al. 2022). Finally, consideration should be given to the needs of special populations, such as those with impaired balance or limited vision. Most head-mounted VR systems are not recommended by the manufacturers for children below the age of 12 or 13.

Because headsets will likely be worn by many different users, maintaining the hygienic standards of equipment also deserves some forethought. High-end solutions include sanitizing systems that are specialized for headsets and that rely on short wave ultraviolet light (UV-C). A less expensive approach involves replacing foam facial interfaces with silicone or vinyl ones that can easily be cleaned with a disinfectant wipe. Disposable face covers or breathable, washable fabric ones are also available on the market.

## 3.2 Choosing a Suitable HMD

> **What to Expect from Off-the-Shelf Head-Mounted VR Hardware**
>
> It goes without saying that all head-mounted VR systems involve some sort of head-mounted display. Some systems, such as the Varjo XR3, which implements hand tracking and inside-out positional tracking, do not require additional hardware. Other systems include hand controllers and base stations or other forms of positional trackers. Many VR systems require tethering to a PC or gaming laptop, such as the HP Fury. Tethering can be accomplished via a physical cable or in some cases, wirelessly through after-market adapters. Some systems, such as the Meta Quest series, are standalone – that is, they operate without external computing support.

> **Space Requirements**
> The amount of physical play space needed for head-mounted VR depends on the kind of interaction that is expected. Viewing a 360° video or experience a simple scene, such as riding a cart on a track, might be suitable for seated participants. Likewise, it may be possible to navigate large virtual environments while walking on a treadmill or sitting in a chair and using the teleport function. On the other hand, if natural walking or other forms of locomotion are required of participants, then a room-scale VR setup with at least 6.5 by 5 feet of open space free of obstacles and hazards is advised.
>
> **Integrated Sensors**
> Off-the-shelf systems support positional tracking of the headset and controllers. Other integrated sensing capabilities vary widely from model to model, ranging from hand and body tracking, eye gaze tracking, facial expression capture, and monitoring of brain and cardiac activities. It should be noted that in some cases, licenses or specialized APIs are necessary to access these data.
>
> **Requisite Technical Expertise**
> Most off-the-shelf systems are designed to work right of the box, requiring only basic technical abilities. Further, some platforms offer game editors that allow simple VR experiences to be created without coding. On the other hand, scripting and other operations necessary to support complex interactions require advance knowledge of the game engine's native programming language. Additionally, more advanced computational skills and knowledge may be required to visualize data from peripheral sensors during engagement in VR, generate event markers in real time, and synchronize and record data streams with event markers.

A central distinction to consider in selecting a VR system hinges around tethered versus untethered models (Table 1). A tethered system is an HMD that is connected physically to a PC through a cord – either USB-c, USB-3, HDMI, or a combination of the 2 (USB, HDMI). It uses the PC that it is connected to for computation, data, power, and rendering. In some cases, a wireless adapter kit may be added to eliminate the need for a physical cable-to-PC connection – but the PC is still required for computation and rendering. On the other hand, untethered systems – often referred to as "standalone" – are endowed with their own fully embedded processor, communications, OS, and power. These devices can range from simple cellphone-based 360° viewers, to the modern gaming headsets of the Meta Quest series and Vive Focus, which feature 6 DoF (degrees of freedom), inside-out tracking, and more. These systems provide convenience for novice VR users and serve well as prototyping devices for people on a budget.

Many HMDs feature integrated sensors above and beyond standard IMUs that track head position and orientation. Infrared eye-tracking, for instance, can support analysis of eye movement, gaze, and pupillary response (PR) of individuals in VR. The HTC Vive Pro Eye series, HP Reverb G2 Omnicept edition (https://www.hp.com/), and Varjo series (https://varjo.com/), among others, all feature integrated eye trackers. Additionally, after-market hardware, such as the Pupil Core from Pupil Labs (https://pupil-labs.com/), is compatible with some VR devices, such as the Vive.

**Table 1** Pros and cons of tethered versus untethered head-mounted VR

| Tethered pros | Tethered cons |
|---|---|
| • Usually guaranteed high performance<br>• Greater FOV (110°)<br>• Higher resolution (depending on PC rig) – supporting ultra high-resolution displays<br>• Higher throughput for multi-threaded work<br>• Customizable for external peripheries (Wi-Fi, hardline connections) and eye-tracking (90+ Hz)<br>• Faster internet speed potential through fiber optics<br>• Usually a high-performance Rig (HPR) can support multiple types of HMD systems and makers<br>• The tether can serve as a boundary or safety line to control distance<br>• No battery to recharge | • Costly setup<br>• A lot of equipment<br>• Not easily portable<br>• Tether can be cumbersome in some experiences |
| **Untethered pros** | **Untethered cons** |
| • Cheaper<br>• Significantly more portable<br>• Typically lighter weight and easier to wear<br>• Faster to deploy for user testing | • Less power and performance<br>• Not as much capability for peripheral or external device integration<br>• Narrow field of view (avg 90° FOV)<br>• Lower resolution<br>• Needs regularly recharging<br>• Often comes with unique user login and side support registration apps or user agreements<br>• Slower Wi-Fi connection |

Current head-mounted VR systems – particularly high-end ones – also tend to support face, body, or hand tracking through integrated cameras – or compatible aftermarket add-ons. These capabilities allow users' facial expressions, as well as the position and orientation of their hands, fingers, and body to be tracked and modeled in the virtual world. Intriguingly, some systems, such as the HP Reverb G2 Omnicept, combine input derived from multiple sensors simultaneously, including face cameras, eye trackers, and a PPG sensor in the headset in order to compute online estimates of a user's cognitive load (Siegel et al. 2021).

Wireless, dry EEG sensors have also been directly incorporated into VR HMDs. The nGoggle (Fig. 2), for instance, features a bank of electrodes over the occipital area of the head that can measure EEG activities associated with visual processing of stimuli presented in the headset (Nakanishi et al. 2017). The DSI-VR300 (https://wearablesensing.com/dsi-vr300/), which is commercially available, offers active electrodes and a somewhat broader distribution of electrodes over parietal and frontal portions of the head in order to support embedded real-time P300 analysis in brain–computer interface (BCI) applications.

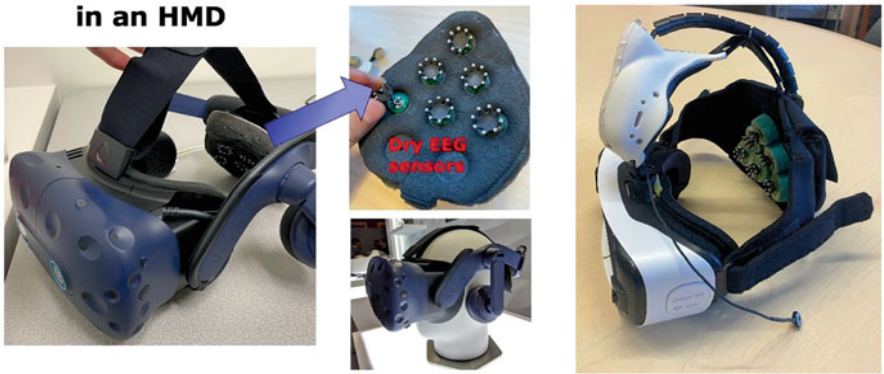

**Fig. 2** Customized VR (HTC Vive, left and nGoggle, right) with integrated EEG electrodes

**Table 2** Comparing custom builds with add-on sensors to pre-built VR sensing packages

| Advantages to custom builds | Advantages to pre-built systems |
| --- | --- |
| • Modular setups assure ideal fit for any subject. Pre-built setups are usually tailored only to average size adults<br>• More flexibility in optimizing sensor positions, sampling rates, active versus passive electrodes, wet versus dry electrodes, and more<br>• Individual parameters can be easily modified for new experiments (e.g., switching from an 8- to a 32-channel EEG system while keeping the same VR hardware or upgrading the VR headset to a different one that has built-in eye-tracking) | • Overall cost may be lower because sensors are integrated in the VR headset<br>• Proprietary data protocols may facilitate (but may also hinder) access to the data<br>• Integrated sensors may be more comfortable given the snug fit that is required of the headsets for an immersive VR experience |

We feel it is important to mention, however, that EEG, PPG, ECG, and other physiological signals have been successfully recorded by the first and second authors of this chapter during engagement in VR-based paradigms using add-on rather than integrated sensors. Indeed, if done correctly, add-ons can provide a more suitable and higher performing solution than currently available pre-built setups because add-on systems are typically optimized for capturing data tailored to specific research needs, whereas pre-built systems typically prioritize user experience and ease of implementation. Table 2 outlines the advantages of both of these approaches. Currently, a variety of mobile EEG options are available, including the Brain Systems Live Amp (https://www.brainproducts.com/), the mBrainTrain Smarting Pro (https://mbraintrain.com/), and the versatile, compact Mentalab ExplorePlus (https://mentalab.com/). It is worth noting that the ExplorePlus is currently the world's smallest research grade system (Niso et al. 2023) and flexibly supports both wet and dry EEG or ECG in customized configurations. For more information on other forms of neuroimaging, consider chapter "Monitoring Brain Activity in VR: EEG and Neuroimaging" in this volume.

As a precautionary note, one of the primary drawbacks of building a VR research system using consumer-grade hardware stems from the risk of obsolescence. The researcher is subject to shifting corporate inclinations. As of writing this chapter, the HTC Vive Pro Eye has already been discontinued. While the upcoming release of an HTC Vive Pro 2 Eye has been promised, the eye-tracking capabilities of this second-generation system may not be identical to the original version. Likewise, announcements have unfortunately been made of plans to phase out the HP Reverb G2 Omnicept edition altogether.

## 3.3 Eye-tracking in VR

With the growing availability of HMDs that feature mobile eye trackers, new possibilities have opened for studying naturalistic, unconstrained gaze behaviors in 3D space (see also chapter "Eye-Tracking in VR" for an in-depth discussion about eye-tracking and head tracking in virtual environments). Through continuous tracking of data about the user's head location and orientation within the virtual world, the head-mounted system can establish the distance between the user and objects and other features of the virtual world at all times, making it possible to compute gaze points as 3D vectors using relative eye position co-ordinates. A gaze intersection point (GIP) is the nearest point in the virtual environment that is intersected by the vector-based gaze ray. Thus, in virtual paradigms, it is possible to study eye movements and gaze both through traditional approaches that estimate changes in visual angle and point of regard based on information about the relative position of the eye in the head, as well as through the analysis of GIPs.

Ongoing work in the primary author's lab suggests that caution must be exercised when analyzing eye movement in contexts that also involve unconstrained head movement, as vestibular-ocular reflex (VOR) can confound the differentiation of fixations from saccades. VOR triggers compensatory eye movements that allow an individual to maintain gaze on a fixed object while moving his or her head. During head movements that trigger VOR, the rate of change of the position of the eye-in-head during a fixation can actually exceed the rate of change of eye position that occurs during small saccades (Tatler et al. 2019), making it extremely difficult to discern the onset of a fixation using algorithms that rely on pupil movement velocity. In other words, even though a person's gaze may have already reached a target, the persistence of low amplitude eye movement may hinder precise determination of fixation onset.

This possibility is supported by Fig. 3, which plots fixation-related potentials (FRPs) derived over CPz in a classic P300 oddball paradigm involving either a saccade only (left) or a saccade and a head turn (right) to the standard or deviant target. In the saccade only condition, fixation-locking yields a robust P300 effect

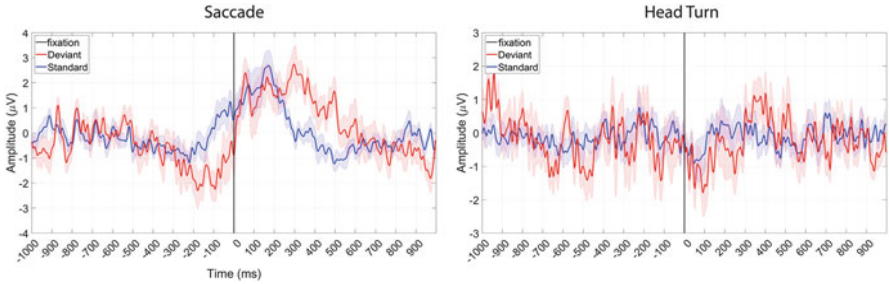

**Fig. 3** Fixation-related potentials computed in response to high (red) or low (blue) probability visual targets. Participants made a saccade to the target (left) or a head turn and a saccade (right). Time zero is fixation onset

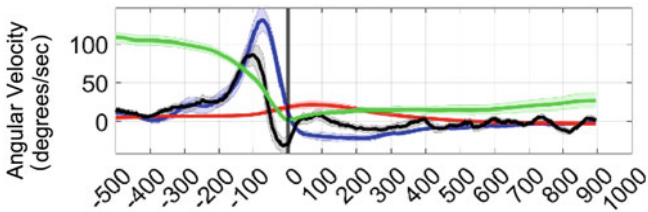

**Fig. 4** Angular velocity of head (red) and eye (blue) movement during a head turn and saccade to a peripheral target. Gaze velocity is plotted in black. The green line indicates the rate of change of the distance of gaze points from the target

from approximately 250 to 550 ms post-fixation. It also reveals sensitivity to standards versus deviants before fixation onset. On the other hand, in the head turning condition, the oddball response is smaller and more difficult to distinguish from the response to standards, and no reliable effects before fixation onset can be discerned. These mushier results in the head turning condition can likely be attributed to two primary factors – namely, imprecise fixation-locking due to VOR, as well as muscle artifacts introduced into the raw EEG during head movement (Wu et al. 2023).

As an alternative to classifying fixations solely on the basis of pupil velocity, it may be possible to use information from the GIP stream to understand when a fixation is occurring on a target of interest. Algorithms can be developed that detect when the GIP is in contact with a target irrespective of pupil movement velocity. Figure 4, for instance, plots the angular velocity of head movement (red line), eye movement (blue line), and gaze (black line), as well as the rate of change in distance of each GIP co-ordinate relative to the target (green line), during a head turn to a peripheral visual target. Time zero is the moment when the GIP first overlaps with the target (causing the distance from the target to register as zero). This time point is presumably the onset of the fixation on the target. Notably, while the rate of change of gaze approaches zero at this moment, both the head and the eyes appear to continue to move in opposite directions.

## 3.4 Synchronizing Multimodal Signals and Recordings of Game Play

If a research paradigm demands heterogeneous data streams – such as continuous physiological measurements coupled with eye-tracking and motion capture data – the challenge of aligning these diverse time series in a common temporal frame is non-trivial. The primary author's research group uses Lab Streaming Layer (LSL) – a freely available API that supports the streaming and recording of multimodal data over a local network and integration of these data streams with event codes generated by a stimulus presentation package (Ojeda et al. 2014). In traditional experiments, stimulus presentation and data acquisition are usually managed by independent systems. However, in VR-based paradigms, the virtual world serves as the "stimulus" and in the case of Unity applications, the VR system can stream event codes related to events either generated internally by the game or effected by the user. Additionally, the VR system can stream physiological and behavioral data obtained from its own built-in sensors (e.g., from the HMD head tracker, from other peripheral IMU trackers, from eye-tracking systems, and so forth) to be synchronized via LSL with time-series data obtained from other sensors that are not integrated with the VR system.

Examples of event codes streamed to LSL by the authors include trigger pulls to pick up virtual objects, ratings of flow during an in-game survey, and the names of virtual objects fixated during game play. Many other events could be created based on participants' location in and interaction with the virtual environment – including body movements, eye movement, and speech. Through a dynamic library file available in the LSL package, it is possible to define streams associated with these event types and then specify updating functions that push event information to the stream at regular intervals. Each stream is recorded as a vector of values (e.g., event codes) and associated timestamps. Because LSL allows heterogeneous data streams to be aligned within a common time scale, it is possible to synchronize event streams during offline analysis with other simultaneously recorded physiological and behavioral data acquired during the experiment. This approach can support event-related analyses, including event-related potentials (ERPs), FRPs, task-evoked pupillary response (TEPR), and more.

To relate time-series data to the player's experience in the virtual world, it may be useful to record not only event markers, but also a video capture of the game play (see Sect. 3.5). A fairly simple approach to synchronizing a screen capture with the other data streams involves creating a visible marker on the screen that is activated by a recordable event. For instance, a flash of color could appear in the corner in response to a trigger pull, that is recorded as a separate LSL stream. In this way, one can align various time-series data recordings with the known time of the trigger pull and the known time of the color flash in the screen capture.

## 3.5 Capturing Video

Video capture of a VR session falls into 3 unique categories – namely, traditional screen capture, virtual camera capture, and mixed reality capture. Traditional screen capture involves recording the scene that is cast to the flat 2d desktop monitor. It typically reflects the user's egocentric perspective. Screen capture tools have become ubiquitous and are deployed on almost all operating systems (e.g., Quicktime on Mac, Win+G on Windows). Many third-party apps are also available such as Nvidia GeForce Experience (https://www.nvidia.com/enus/geforce/geforce-experience/) and Open Broadcast Software (OBS) (https://obsproject.com/) – an open-source video audio mixer software for parsing content from multiple streams into a single directed format that can then be recorded or streamed in real time to external sources. This platform would be useful in cases where you may have multiple virtual cameras and want to stitch them all together in real-time.

Alternatively, if one wishes to capture activities occurring at a specific location in the virtual environment or from an allocentric point of view, virtual cameras can be placed into the game programmatically and operated as viewports into the game. With them you can follow the player, free flying (as if controlling a drone). Static cameras can also be stationed in the 3D world space, making it possible to capture a 360° view from a fixed location. This technique can also be used to replay the point of view of the user and can contain different information from scene casting to the desktop monitor. Virtual cameras have become mainstays for broadcasting in-game events to the outside world and are supported by most game engines. Third-party plugins can be obtained as well.

The last method to be examined in this section is mixed reality capture. It involves capturing the physical, real-world body of the player mixed within the digital world. This technique uses a combination of virtual cameras and real-world cameras calibrated together to place the real person within the confines of the game he or she is playing. Traditional green screen photography and virtual cameras are used to mix worlds. Pervasive in Hollywood filming today, mixed reality capture is more resource intensive than the previously described methods and generally requires a specialized space big enough to mask the backgrounds out and additional equipment to manage the virtual camera interface with a physical device. There are many tools and systems that can now be used to this end, including MixCast (https://mixcast.me/) by Intel and Reality Mixer, which can be obtained through Apple as an iOS-based app that relies on iPads or iPhones (though a beta version that interfaces with certain VR headsets is also available through Steam). Microsoft Mixed Reality Capture Studios is another resource, supporting volumetric video and hologram creation. It is built in natively to the Microsoft Mesh platform and works both with VR and Hololens2.

## 4 Conclusion

This chapter underscores the fact that accomplishing behavioral neuroscience in VR is a cross-disciplinary team effort that mixes the roles of experimentalists, developers, and engineers. However, VR is a versatile medium that can be adapted to different research objectives and the capabilities of different research teams. Off-the-shelf solutions exist in the form of plug-and-play head-mounted systems with integrated sensors, as well as asset libraries, game engines, and customizable game platforms. Alternatively, it is possible to develop specialized content from scratch and modify existing, commercially available VR hardware to suit specific experimental needs.

As outlined in the first section of this chapter, developing a VR-based research platform requires attention to the kinds of interactivity and immersive sensory content – such as visual, auditory, haptic, and olfactory stimulation – that will be put to play in the virtual experience that is part of the research paradigm. As outlined in the second section, one's research goals can guide the selection of hardware and possibly add-on sensors. Further, a variety of open-source solutions can be implemented to overcome the challenge of capturing and synchronizing diverse streams of data. As a final note, advances in signal processing methods and theoretical frameworks for interpreting experimental outcomes are still catching up with the pace of research possibilities opened by VR technologies. The reader is invited to explore the other chapters of this book to gain further insight into solutions that have already been successfully implemented in response to these challenges.

**Acknowledgment** This chapter was supported by grant #IIS-2017042 from the National Science Foundation and and #W911NF-21-2-0126 from the Army Research Laboratory to the first author, and by a grant from the Royal Society of New Zealand Marsden fund (VUW-2005) which supported the second author.